\documentstyle[preprint,aps,eqsecnum,epsf,floats]{revtex}
\begin{document}
\tighten

\def\bfl{{\bbox \ell}}
\def\bull{\vrule height .9ex width .8ex depth -.1ex}
\def\MeV{{\rm MeV}}
\def\GeV{{\rm GeV}}
\def\Tr{{\rm Tr\,}}
\def\nrcpt{NR\raise.4ex\hbox{$\chi$}PT\ }
\def\ket#1{\vert#1\rangle}
\def\bra#1{\langle#1\vert}
\def\ltap{\ \raise.3ex\hbox{$<$\kern-.75em\lower1ex\hbox{$\sim$}}\ }
\def\gtap{\ \raise.3ex\hbox{$>$\kern-.75em\lower1ex\hbox{$\sim$}}\ }
\def\abs#1{\left| #1\right|}
\def\CA{{\cal A}}
\def\CC{{\cal C}}
\def\CD{{\cal D}}
\def\CE{{\cal E}}
\def\CL{{\cal L}}
\def\CO{{\cal O}}
\def\CZ{{\cal Z}}
\def\bvert{\Bigl\vert\Bigr.}
\def\pds{{\it PDS}\ }
\def\ms{MS}
\def\ddq{{{\rm d}^dq \over (2\pi)^d}\,}
\def\ddqm{{{\rm d}^{d-1}{\bf q} \over (2\pi)^{d-1}}\,}
\def\bfq{{\bf q}}
\def\bfk{{\bf k}}
\def\bfp{{\bf p}}
\def\bfpp{{\bf p '}}
\def\bfr{{\bf r}}
\def\dtr{{\rm d}^3\bfr\,}
\def\bfx{{\bf x}}
\def\dtx{{\rm d}^3\bfx\,}
\def\dfx{{\rm d}^4 x\,}
\def\bfy{{\bf y}}
\def\dty{{\rm d}^3\bfy\,}
\def\dfy{{\rm d}^4 y\,}
\def\dfq{{{\rm d}^4 q\over (2\pi)^4}\,}
\def\dfk{{{\rm d}^4 k\over (2\pi)^4}\,}
\def\dfl{{{\rm d}^4 \ell\over (2\pi)^4}\,}
\def\dtq{{{\rm d}^3 {\bf q}\over (2\pi)^3}\,}
\def\dtk{{{\rm d}^3 {\bf k}\over (2\pi)^3}\,}
\def\dtl{{{\rm d}^3 {\bfl}\over (2\pi)^3}\,}
\def\dt{{\rm d}t\,}
\def\frac#1#2{{\textstyle{#1\over#2}}}
\def\darr#1{\raise1.5ex\hbox{$\leftrightarrow$}\mkern-16.5mu #1}
\def\){\right)}
\def\({\left( }
\def\]{\right] }
\def\[{\left[ }
\def\si{{}^1\kern-.14em S_0}
\def\siii{{}^3\kern-.14em S_1}
\def\diii{{}^3\kern-.14em D_1}
\def\p0iii{{}^3\kern-.14em P_0}
\def\fm{{\rm\ fm}}
\def\MeV{{\rm\ MeV}}
\def\CA{{\cal A}}
\def\Czzm{ {\cal A}_{-1[00]} }
\def\Cttm{{\cal A}_{-1[22]} }
\def\Ctzm{{\cal A}_{-1[20]} }
\def\Cztm{ {\cal A}_{-1[02]} }
\def\Czzz{{\cal A}_{0[00]} }
\def\Cttz{ {\cal A}_{0[22]} }
\def\Ctzz{{\cal A}_{0[20]} }
\def\Cztz{{\cal A}_{0[02]} }


\def\spzz{ {Y_{sp0}^{(0)} }}
\def\spzzB{ {Z_{sp0}^{(0)} }}
\def\spzo{ {Y_{sp0}^{(1)} }}
\def\spzt{ {Y_{sp0}^{(2)} }}
\def\ppspz{ {y^{sp0}_0 }}
\def\ppspt{ {y^{sp0}_2 }}
\def\ppspB{ {z^{sp0}_2 }}

\def\Ames{ A }  

\newcommand{\eqn}[1]{\label{eq:#1}}
\newcommand{\refeq}[1]{(\ref{eq:#1})}
\newcommand{\eq}{eq.~\refeq}
\newcommand{\eqs}[2]{eqs.~(\ref{eq:#1}-\ref{eq:#2})}
\newcommand{\eqsii}[2]{eqs.~(\ref{eq:#1}, \ref{eq:#2})}
\newcommand{\Eq}{Eq.~\refeq}
\newcommand{\Eqs}{Eqs.~\refeq}

\def\Journal#1#2#3#4{{#1} {\bf #2}, #3 (#4)}

\def\NCA{\em Nuovo Cimento}
\def\NIM{\em Nucl. Instrum. Methods}
\def\NIMA{{\em Nucl. Instrum. Methods} A}
\def\NPB{{\em Nucl. Phys.} B}
\def\NPA{{\em Nucl. Phys.} A}
\def\PLB{{\em Phys. Lett.}  B}
\def\PRL{\em Phys. Rev. Lett.}
\def\PRD{{\em Phys. Rev.} D}
\def\PRC{{\em Phys. Rev.} C}
\def\PRA{{\em Phys. Rev.} A}
\def\ZPC{{\em Z. Phys.} C}
\def\PREP{{\em Phys. Rep.}  }
\def\ANN{{\em Ann. Phys.} }
\def\SCI{{\em Science} }

\preprint{\vbox{
\hbox{ NT@UW-98-17}
\hbox{ DUKE-TH-98-166}
}}
\bigskip
\bigskip

\title{Parity Violation in Effective Field Theory
and the Deuteron Anapole Moment\footnote{
This paper is published in 
{\it Nucl. Phys.} {\bf A644} 235-244 (1998).
The last section contains an erratum and addendum to the 
published version.}
}
\author{Martin J. Savage and Roxanne P. Springer\footnote{\rm  
On leave from the Department of Physics,   Duke University, Durham NC 27708.
    \ \tt rps@redhook.phys.washington.edu.}}
\address{Department of Physics, University of Washington, Seattle, WA 98195
  \\ {savage@phys.washington.edu, rps@redhook.phys.washington.edu}}
\maketitle

\begin{abstract}
  Effective field theory, including pions,
  provides a consistent and systematic
  description of nucleon-nucleon strong interactions up to
  center-of-mass momentum ${\bf p} \sim 300\ {\rm MeV}$ per nucleon.
  We describe the inclusion of hadronic parity violation
  into this effective field theory
  and find an analytic form for the deuteron anapole moment
  at leading order in the expansion.
\end{abstract}

\section{Introduction}

An increasing amount of both experimental and theoretical
interest is being focused on nuclear parity violation.
The problems in reproducing parity violating observables in
light nuclei (such as $^{18}$F\cite{Fluorine}) and in the nuclear 
anapole measurement of cesium\cite{anawei}
by parameters in
single boson exchange potentials\cite{PVprobs,PVprobsb}
suggest that we re-examine the theoretical framework with which 
hadronic parity violation is studied.
The low momentum transfers involved in these parity violating  
processes makes them amenable to an effective field theory treatment.
After much effort
\cite{Weinberg1,KoMany,Parka,KSWa,CoKoM,DBK,cohena,Fria,Sa96,LMa,GPLa,Adhik,RBMa,Bvk,aleph,Parkb,Gegelia,steelea,KSW},
a consistent power counting and calculational scheme has
been developed which provides a systematic treatment of
the low-momentum nucleon-nucleon strong interactions~\cite{KSW}, up
to a center-of-mass momentum for each nucleon of
${\bf p}~\sim~300\ {\rm MeV}$.
Analytic expressions for $NN$ scattering phase shifts and 
deuteron observables~\cite{KSW2,CGSS}
have been obtained which describe the data well.
Using this technique we construct an effective field theory 
describing weak
interactions in the two-nucleon sector, and as an example 
obtain an analytic expression for
the deuteron anapole moment at leading order in the expansion.

\section{Strong Interaction Amplitudes}

The strong interactions of the pions are described by the Lagrange density
\begin{eqnarray}
{\cal L}_0 & = &  {f^2\over 8} Tr D_\mu \Sigma
D^\mu \Sigma^\dagger
+ {f^2\over 4} \lambda Tr m_q (\Sigma + \Sigma^\dagger)
\ +\ ...
\ \ \ ,
\end{eqnarray}
where the pion fields are incorporated
in a special unitary matrix,
\begin{equation}
\Sigma = \exp {2i\Pi\over f},\qquad \Pi =
\left(\begin{array}{cc}
\pi^0/\sqrt{2} & \pi^+\\ \pi^- & -\pi^0/\sqrt{2}\end{array} \right),
\end{equation}
with $f=132\ \MeV$.
The strong interactions of the nucleons are described by a Lagrange density
that we write as
\begin{equation}
{\cal L} = {\cal L}_0 + {\cal L}_1 + {\cal L}_2 + \ldots,
\end{equation}
where ${\cal L}_n$ contains $n$-body nucleon operators.

The one-body terms in the Lagrange density are
\begin{eqnarray}
{\cal L}_1 & = &  N^\dagger \left(i D_0 + {{\bf D}^2\over 2M}\right) N 
+ {ig_A\over 2} N^\dagger {\bf \sigma} \cdot 
(\xi {\bf D} \xi^\dagger - \xi^{\dagger} {\bf D} \xi)N
\ +\ ...
\label{eq:lagone}
\ \ \ \ ,
\end{eqnarray}
where $g_A = +1.25$ and
the ellipses denote operators with more insertions of the light quark
mass matrix and more spatial derivatives.
The two-body Lagrange density contributing to  S-wave 
interactions may be written as
\begin{eqnarray}
{\cal L}_2 &=& -\(C^{(\siii)}_0+ D^{(\siii)}_2 \lambda\Tr m_q\))
(N^T P_{i,0} N)^\dagger(N^T P_{i,0} N)
\nonumber\\
 & + & {C_2^{(\siii)}\over 8}
\left[(N^T P_{i,0} N)^\dagger
\left(N^T \left[ P_{i,0} \overrightarrow {\bf D}^2 +\overleftarrow {\bf D}^2 P_{i,0}
    - 2 \overleftarrow {\bf D} P_{i,0} \overrightarrow {\bf D} \right] N\right)
 +  h.c.\right]
\nonumber\\
&  & -\(C^{(\si)}_0+ D^{(\si)}_2 \lambda\Tr m_q\)) (N^T  P_{0,a}
N)^\dagger(N^T P_{0,a} N)
\nonumber\\
 & + & {C_2^{(\si)}\over 8}
\left[(N^T P_{0,a} N)^\dagger
\left(N^T \left[ P_{0,a} \overrightarrow {\bf D}^2 +\overleftarrow {\bf D}^2 P_{0,a}
    - 2 \overleftarrow {\bf D} P_{0,a} \overrightarrow {\bf D} \right] N\right)
 +  h.c.\right]
\nonumber\\
 & + & ...
\ \ \ \ ,
\label{eq:lagtwo}
\end{eqnarray}
where the ellipses denote operators involving more insertions of the light
quark mass matrix, meson fields, and spatial derivatives.
The $P_{i,0}$ and $P_{0,a}$ are spin-isospin projectors defined by
\begin{eqnarray}
P_{i,0} & \equiv & {1\over\sqrt{ 8} } \sigma_2\sigma_i\tau_2\ , 
\qquad \Tr P_{i,0}^\dagger
P_{j,0} = {1\over 2} \delta_{ij}\ ,
\nonumber\\
P_{0,a} & \equiv & {1\over\sqrt{ 8} } \sigma_2\tau_2\tau_a\ , 
\qquad \Tr P_{0,a}^\dagger
P_{0,b} = {1\over 2} \delta_{ab}
\ \ \ .
\end{eqnarray}
The coefficients have been determined from fits to the S-wave phase shifts in
both the $\si$ and $\siii$ channels~\cite{KSW}.  Renormalized at $\mu=m_\pi$
in the power divergence subtraction scheme (PDS)\cite{KSW},
it is found that 
\begin{eqnarray}
C_0^{(\si)} & = & -3.34\fm^2
\ ,\qquad
D_2^{(\si)} =-0.42\fm^4
\ ,\qquad
C_2^{(\si)} =3.24\fm^4
\ \ \  ,
\nonumber\\
C_0^{(\siii)} & = & -5.51\fm^2\ ,\qquad 
D_2^{(\siii)} =1.32\fm^4\ ,\qquad 
C_2^{(\siii)} =9.91\fm^4
\ \ \  .
\end{eqnarray}

The relative importance of an operator to a particular
observable is given by the running of the operator in  the
infrared.  Dimensional regularization with the PDS scheme provides 
a  simple and consistent determination of this running.
Consider a general operator arising from strong interactions 
that connects states with angular momentum $L$ and $L^\prime$.
If this operator has $m$ insertions of the light
quark mass matrix $m_q$ and $2d = L+L^\prime + 2n$ spatial gradients,
we will denote its coefficient  by $C^{(L,L^\prime)}_{m,n}$.
It was shown in \cite{KSW,KSW2} that such a coefficient scales like
$\mu^{-(n+m+1)}$ for the $\si$ and  $\siii-\diii$ partial waves, and like
$\mu^0$ for all other partial waves.
For nucleon nucleon scattering in the $\si$ channel 
the momentum independent contact term with no insertions of $m_q$ has
coefficient $C_0^{(\si)} = C_{0,0}^{(0,0)}\sim \mu^{-1}$, 
and  is the leading contribution.
Each loop graph in the time-ordered product of two insertions of this operator
scales like $\mu C_{0,0}^{(0,0)}\sim 1$.  Therefore the bubble chain of such
insertions constitutes the leading amplitude.
At subleading order two additional operators appear.
There is a contribution from a single insertion of an
operator with $d=1,m=0$ (coefficient $C_2^{(\si)}$),
and another from a single insertion of an operator
with $d=0,m=1$ (coefficient $D_2^{(\si)}$).
At the same order, 
there are contributions from the exchange of a single potential
pion, which scales like $\mu^0$.
For the remaining discussion we will represent 
the small expansion parameters, the external momentum ${\bf p}$, 
the quark masses $m_q$, and the renormalization
scale $\mu$, by the generic quantity $Q$.

\section{Parity Violating Interactions}

The standard model of electroweak interactions gives rise to parity violating
four-quark operators that exhibit enhanced symmetry in particular limits.
For instance, in the limit that the angles of the Cabibbo-Kobayashi-Maskawa (CKM)
matrix and the weak mixing angle vanish, and the mass of the two
quarks in each generation are degenerate, the standard model Lagrange density
has a global $SU(2)_L\otimes SU(2)_R$ chiral symmetry\cite{DSLS}.
All parity violation in this limit must be $\Delta I=0$.  
The large difference between the mass of the charm  and strange quarks 
breaks this symmetry.
For hadronic processes the strange quarks can be considered
dynamical while the charm quarks are ``integrated out'' of the theory.
Therefore, in the limit of vanishing mixing angles in the CKM matrix and in the
neutral gauge boson sector, all $\Delta I=1$ parity violation comes from
strange quarks.  In the real world with non-zero mixing angles, the up and down
quarks also contribute, but with coefficients suppressed by  
$\sin^2\theta_w$, and by the Cabibbo angle.

Excellent reviews of  parity violating observables in the context of
meson exchange models of weak nucleon-nucleon scattering
appear in \cite{adelhax,DDH,MMa,DZ,HHpv}.
In this work we do not reproduce the results obtained in
potential models for various parity violating observables, but 
instead focus on
the effective field theory description of parity violation.
This will facilitate the systematic inclusion of higher order effects 
such as relativistic corrections and  dynamical meson exchange.
A first step  was undertaken in \cite{KSa} where a Lagrange density 
consistent with chiral symmetry was constructed for the 
single nucleon sector.
As with any effective field theory
constructed to reproduce the low-energy behavior of 
some unknown or unsolved theory, there are couplings in the theory
that must be determined by experiment 
in order for the theory to be  predictive.

The parity violating four-quark operators generated by the standard model of
electroweak interactions can be classified according to how they transform under
$SU(2)_L\otimes SU(2)_R$ chiral transformations\cite{KSa}.
By explicit construction they have decompositions
$(1,1),(3,1)\oplus(1,3)$, and $(5,1)\oplus(1,5)$, and so can be unambiguosly
classified by their transformation under the isospin subgroup.
Operators transforming as $(3,3)$  also occur but do not violate parity.
The following operators\cite{KSa}
\begin{eqnarray}
  X_L^a & = & \xi^\dagger\tau^a\xi
  \ \ \ ,\ \ \ 
 X_R^a = \xi\tau^a\xi^\dagger
 \nonumber\\
  X_L^a -  X_R^a & = & -{2\sqrt{2}\over
    f}\epsilon^{a\alpha\beta}\pi^\alpha\tau^\beta\ +\ ...
  \ \ ,\ \
  X_L^a +  X_R^a \ =\ 2\tau^a\ +\ ...
\ \ \ ,
\end{eqnarray}
transform as $X\rightarrow U X U^\dagger$ under chiral transformations.
Rewriting the Lagrange density of \cite{KSa} in terms of two-component
non-relativistic nucleon fields, so that they are expressed in terms of the 
same field operators as the strong $NN$ Lagrange density in
eqs.(\ref{eq:lagone}) and (\ref{eq:lagtwo}),
we have\footnote{
Note that the sign of $h_{\pi NN}^{(1)}$ is opposite that used in\cite{KSa}.
}
\begin{eqnarray}
  {\cal L}_{\Delta I=0}^{\rm pv} & = &
  h_V^{(0)}  N^\dagger \Ames_0\ N\ +\ ...
  \nonumber\\
  {\cal L}_{\Delta I=1}^{\rm pv} & = &
  \ - h_{\pi NN}^{(1)} {1\over 4}\ f_\pi\ N^\dagger ( X_L^3-X_R^3 ) N
  \nonumber\\
  & + & 
  h_V^{(1)} {1\over 2}  N^\dagger\ N\  Tr\left[ \Ames_0 ( X_L^3+X_R^3 )\right]
    \ -\
    h_A^{(1)} {1\over 2}  N^\dagger\sigma^a\  N\  Tr\left[ {\bf \Ames}_a (
      X_L^3-X_R^3 )\right]
    \ +\ ...
  \nonumber\\
  {\cal L}_{\Delta I=2}^{\rm pv} & = &
   h_V^{(2)}\  {\cal T}_{ab} \ N^\dagger \left( X_R^a\Ames_0 X_R^b + X_L^a\Ames_0
     X_L^b\right) N
   \nonumber\\
   & - & 
   h_A^{(2)}\  {\cal T}_{ab}\  N^\dagger \left( X_R^a\  {\bf \sigma}\cdot {\bf
       \Ames}\  X_R^b - X_L^a\  {\bf \sigma}\cdot {\bf \Ames}\  X_L^b\right) N
   \ +\ ...
   \ \ \ \ ,
   \label{eq:weakL}
\end{eqnarray}
where the ellipses represent higher order terms  
in the chiral and momentum expansion.
$\Ames_\mu$ is the axial vector meson field 
$\Ames_\mu = {i\over 2}
\left( \xi^\dagger D_\mu \xi\ -\ \xi D_\mu\xi^\dagger\right)$,
with $D_\mu = \partial_\mu + i e A_\mu$, 
and ${\cal T}_{ab}$ is defined in \cite{KSa}\ to be
\begin{eqnarray}
  {\cal T}_{ab} & = & {1\over 3}\left(\matrix{1&0&0\cr 0&1&0\cr 0&0&-2}\right)
  \ \ \ \ .
\end{eqnarray}
In the power counting we employ, the size of the coefficients in the Lagrange
density
of eq.~(\ref{eq:weakL})\ ($h_{\pi NN}^{(1)},  h_V^{(0,1,2)}$, and $h_A^{(1,2)}$)
is set by naive dimensional analysis (NDA).
Were the coefficients to dramatically deviate from their NDA size the 
power counting must be modified.

Expanding the  interactions in eq.(\ref{eq:weakL}) to ${\cal O}(\pi)$ 
gives
\begin{eqnarray}\label{wope}
  {\cal L}_{\pi NN}^{\rm pv} & = & -i\ h_{\pi NN}^{(1)}\ \pi^+\ p^\dagger n\ +\
  {\rm h.c.}\
  \ \ \ \ ,
\end{eqnarray}
where the $h_V$-type interactions involving a single $\pi$ do not contribute
due to the conservation of the vector current.
The strong interaction between two nucleons induced by the exchange of a single
potential pion scales like $Q^0$ in the effective field theory; 
two factors of momentum from the derivative coupling and two powers
of $q$ and $m_\pi$ in the pion propagator give
\begin{eqnarray}
  V({\bf q}) & \sim &
  { ({\bf \sigma}\cdot {\bf q}) ( {\bf \sigma}\cdot {\bf q})
    \over {\bf q}^2+m_\pi^2}
\ \ \  .
\end{eqnarray}
The weak interaction induced between two nucleons from the  single insertion
of the operator with coefficient $h_{\pi NN}^{(1)}$ scales like $Q^{-1}$, since
the weak $\pi NN$ vertex is independent of momentum,
\begin{eqnarray}
  V({\bf q}) & \sim & { {\bf \sigma}\cdot {\bf q} \over {\bf q}^2+m_\pi^2}
\ \ \  .
\end{eqnarray}
The meson-nucleon vertices induced by the other operators  in
eq.(\ref{eq:weakL}) involve more derivatives so their contribution to the weak
nucleon nucleon interaction is suppressed by additional powers of the expansion
parameter $Q$.
If for some reason $h_{\pi NN}^{(1)}$ is much smaller than indicated by NDA, as
suggested by the $^{18}$F experiments\cite{Fluorine}, while the others 
coefficients remain their NDA size, then this weak one pion exchange interaction
 will no  longer be the leading order.

In addition to the contribution from the exchange of one or more mesons, there
are contributions to the weak interaction between nucleons coming from weak
four-nucleon contact operators.
The leading order weak interaction (CP invariant) between nucleons 
in an S-wave in both initial and final states is described by the 
Lagrange density
\begin{eqnarray}
 {\cal L} & = &  
\eta^{(1)}_{\si\si ;0} \ {f\over 2} \  
 \left(N^T P_{0,a} N\right)^\dagger
 N^T  \{P_{0,a},X_L^3-X_R^3\}_T N 
 \ \ \ +\ \ {\rm h.c.}
 \nonumber\\
 & = &
 -i\  \eta^{(1)}_{\si\si ;0}\ 
 \left( p^T\sigma_2 n + n^T\sigma_2  p\right)^\dagger
 \left( p^T\sigma_2 p\ \pi^- - n^T\sigma_2 n\ \pi^+\right)
 \ +\ {\rm h.c.}\ +\ {\cal O}(\pi^3)
 \ \ \ .
\end{eqnarray}
Notice that this does not generate an interaction involving only two nucleons,
but does give rise to an interaction between two nucleons and pions.
We have defined the modified anti-commutator by
\begin{eqnarray}
  \{A, B\}_T & = & A B + B^T A
  \ \ \ .
\end{eqnarray}
Coefficients of these operators are defined to have the form
$\eta^{(\Delta I)}_{ {^{(2s_i+1)}{L_i}_{J_i}} {^{(2s_f+1)}{L_f}_{J_f}} ;n}$,
where the nucleon quantum numbers are defined to be those in the absence of
pions, and $n$ is the number of $\nabla^2$ and $m_q$ insertions.
The $\beta$-function for these operators shows that
that coefficient $\eta^{(1)}_{\si\si ;0}$  scales like $\mu^{-2}$.
In general, operators in this channel with $n$ insertions of $\nabla^2$ and 
$m$ insertions of the light quark mass matrix scales like 
$\mu^{-2(n+m+1)}$.
At next order there are contributions from operators involving more
insertions of the light quark mass matrix, insertions of the axial vector meson
field $\Ames_\mu$ and more derivatives acting on the nucleon fields.
This is similar to those found in eq.(\ref{eq:lagtwo}).
There is no analogous interaction involving nucleons in the $\siii$ channel 
that is CP invariant.

The dominant contribution to low-energy parity violation in $pp$ scattering
is from the $\si-\p0iii$ weak amplitude interfering with the strong 
scattering amplitude.
The four nucleon operators for such weak scattering are, for $\Delta I=0$
\begin{eqnarray}
  {\cal L} & = & 
\eta_{\si\p0iii ;0}^{(0)}  \left(N^T P_{0,a} N\right)^\dagger
N^T\left[ P_{i,a}\overrightarrow{\bf D}^i  -\overleftarrow{\bf D}^i
  P_{i,a}\right] N\ +\ {\rm h.c.}
\ \ \ ,
\end{eqnarray}
and for $\Delta I=1$
\begin{eqnarray}
  {\cal L} & = & 
\eta_{\si\p0iii ;0}^{(1A)} 
 \left(N^T \{ P_{0,a} ,X_L^3+X_R^3\}_T N\right)^\dagger
N^T\left[ P_{i,a}\overrightarrow{\bf D}^i  -\overleftarrow{\bf D}^i
  P_{i,a}\right] N\ +\ {\rm h.c.}
\nonumber\\
& + &
\eta_{\si\p0iii ;0}^{(1B)}  \left(N^T P_{0,a} N\right)^\dagger
N^T\left[\{ P_{i,a} ,X_L^3+X_R^3\}_T 
\overrightarrow{\bf D}^i  -\overleftarrow{\bf D}^i
  \{ P_{i,a} ,X_L^3+X_R^3\}_T \right] N\ +\ {\rm h.c.}
\ \ \ .
\end{eqnarray}
We have not shown the $\Delta I=2$ Lagrange density.
The coefficients of these operators, $\eta_{\si\p0iii ;0}^{(0)}$,
$\eta_{\si\p0iii ;0}^{(1A)} $, and $\eta_{\si\p0iii ;0}^{(1B)}$ scale like
$\mu^{-1}$ and hence contribute to the $pp$ scattering amplitude at order $Q^0$.
In general, operators in this channel with $n$ insertions of $\nabla^2$ and 
$m$ insertions of the light quark mass matrix scales like 
$\mu^{-(n+m+1)}$.
At next order in the expansion terms involving more insertions of spatial
gradients and the light quark mass matrix give a contribution at order $Q^1$,
suppressed by only one power of $Q$ in the expansion.
Therefore, at next-to-leading order (NLO), there are contributions from both
subleading terms in the strong interaction  and from
subleading terms in the weak interaction.
The construction of the four nucleon operators contributing to weak processes
between higher partial waves follows straightforwardly from the above
discussions, and we will not detail them in this work.


\section{The Anapole Moment of the Deuteron}

As an example of a parity violating quantity that can be described 
using this effective theory, we will calculate the anapole moment of the 
deuteron at leading order in the expansion.
This is of interest because of the possibility of measuring or constraining 
the coefficient $h^{(1)}_{\pi NN}$.
Attempts to determine $h^{(1)}_{\pi NN}$ through parity violating observables 
in nuclei have been frustrated by the uncertainties inherent in 
complicated nuclear systems.
Parity violating observables in the interaction between leptons and 
single nucleons also receive contributions
from $h^{(1)}_{\pi NN}$ but
there are contributions from both isoscalar and isovector
$Z^0$ mediated interactions as well.
In contrast, the deuteron has only isoscalar interactions,
dominated by the strangeness content  of the deuteron.
Fortunately, there are currently two experiments under 
consideration\cite{npprop,sample}
which will look at the theoretically clean deuteron system.
The $\vec n p\rightarrow d\gamma$ process\cite{npprop} has been treated 
in this effective field theory
in ref.~\cite{KSSWpv} and parity violation in electron-deuteron 
scattering\cite{sample} is the subject of this section.

The anapole moment of a particle is its spin dependent coupling to the
electromagnetic field.
For the nucleon
\begin{eqnarray}
  {\cal L} & = & A_N\ {1\over M_N^2}\ N^\dagger\  \sigma^k\  
  N\  \partial_\mu F^{\mu k}
\ \ \ .
\label{eq:nucana}
\end{eqnarray}
The equations of motion for the electromagnetic field,
$\partial_\mu F^{\mu\nu} = e \overline{\Psi}\gamma^\nu \Psi$,
allow this operator to
be written entirely in terms of local four-Fermi contact terms.
It is convenient to keep the operator in the form of
eq.~(\ref{eq:nucana})
for calculational purposes.
Parity violating lepton hadron scattering that depends upon the spin of the
hadronic target receives contributions from both the anapole moment of the
object and from the exchange of $Z^0$ gauge bosons between the lepton current
and the hadronic target.
The anapole moment of elementary particles depends upon the choice of $R_\xi$
gauge\cite{Musolfa}, since its contribution  cannot be separated
from the contributions of $W$-box and $Z$-loop graphs, which all occur
at order $g^4$ in the electroweak coupling.
However, since the anapole moment of the hadrons (composite objects) is
dominated by the long distance behavior of QCD, and the weak vertices at the
hadronic level are independent of the choice of $R_\xi$ gauge, these are
well-defined contributions to the spin dependent weak scattering amplitude.

\begin{figure}[t]
\centerline{\epsfxsize=4.0in \epsfbox{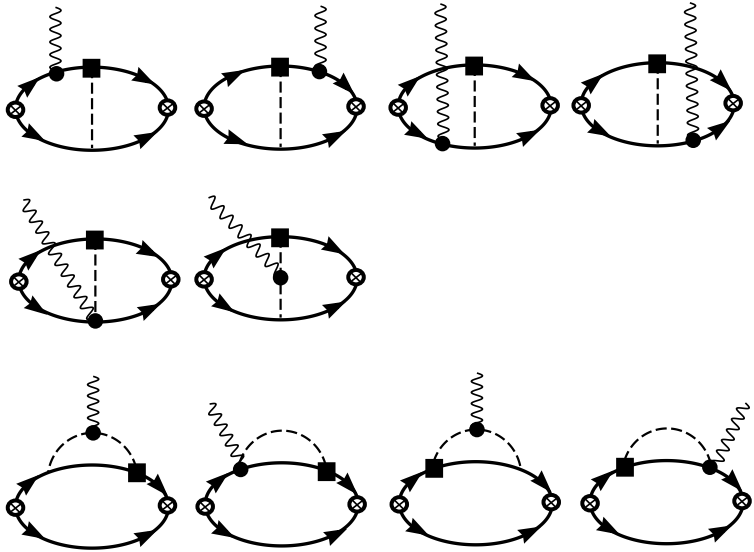}}
\noindent
\caption{\it Leading order diagrams contributing to the deuteron
  anapole moment.  
  The crossed circles denote operators that create  
  or annihilate two nucleons with
  the quantum numbers of the deuteron.
  The solid square denotes the weak operator with 
  coefficient $h^{(1)}_{\pi NN}$ and the solid circle denotes minimal coupling
  to the electromagnetic field.
  Wavy lines are photons, solid lines are nucleons, and dashed lines are mesons.
  }
\label{fig:ana}
\vskip .2in
\end{figure}

The anapole moment of the deuteron is $A_D$, the coefficient in
\begin{eqnarray}
  {\cal L} & = & i A_D\ {1\over M_N^2}\
  \epsilon_{abc} D^{a\dagger} D^b  \partial_\mu F^{\mu c}
  \ \ \ ,
\end{eqnarray}
where  $i \epsilon_{abc} D^{a\dagger} D^b$ is the deuteron spin operator.
The leading contribution to  $A_D$ comes from  weak meson exchange 
starting at order
$Q^{-1}$, dominated by the operator with coefficient 
$h^{(1)}_{\pi NN}$
( see fig.~(\ref{fig:ana}).)
The weak four nucleon
operators  start contributing at order $Q^{0}$ and we will neglect them in this
work.

Explicit computation of the graphs shown in fig.~(\ref{fig:ana}) leads to a
deuteron  anapole moment  of
\begin{eqnarray}
  A_D & = & {e g_A h^{(1)}_{\pi NN} M_N^2 \over 12 \pi f m_\pi} 
  \left[1\ +\  {m_\pi^2 + 3 \gamma m_\pi + 12 \gamma^2\over 6 (m_\pi+2\gamma)^2}
  \right]
\ \ \ ,
\label{eq:Dana}
\end{eqnarray}
where $\gamma = \sqrt{M_N B}$ is the binding momentum of the 
deuteron with $B$ the deuteron binding energy.
The first term in eq.~(\ref{eq:Dana}) is from the
graphs that give the dominant contribution to the nucleon anapole 
moment\cite{Musolfa} (the last four graphs in 
fig.~(\ref{fig:ana})),  and
the second term is the contribution from potential pion exchanges between the
two nucleons in the deuteron
(the first six graphs in  fig.~(\ref{fig:ana})).  
These two terms are the same order in
the expansion, but the contribution from the individual nucleon anapole
moments is numerically larger.  
Higher order corrections to this result that are parametrically suppressed by
the expansion parameter(s) arise from higher dimension operators in both the
weak sector and in the strong sector.
These corrections are typically on the order of $30\%$.
In addition to the contributions from the higher dimension weak meson operators
in (\ref{eq:weakL}) and the contributions from the weak four nucleon operators
there is a contribution from four nucleon anapole operators,
e.g.
\begin{eqnarray}
{\cal L} & = &  
A_{4N}\ i\epsilon_{ijk} (N^T P_{i,0} N)^\dagger(N^T P_{j,0} N)\ \partial_\mu F^{\mu k}
\ +\ ...
\ \ \ \ ,
\label{eq:fourana}
\end{eqnarray}
that contributes at next-to-next-to-leading  order in the $Q$ expansion.
The coefficient $A_{4N}$ scales like $\mu^{-2}$ in the PDS scheme.

The deuteron spin dependent amplitude for electron deuteron scattering 
receives contributions from both the deuteron
anapole moment and from the direct exchange of a $Z^0$ 
(and the associated radiative corrections).
The spin-dependent matrix element for $e D$ scattering, 
at tree-level in the  standard model is
\begin{eqnarray}
  {\cal A}^{(pv)} & = &
  10^{-7} \left( 5.4  g_{\Delta S} \ -\ 0.17 \left( {h^{(1)}_{\pi NN}\over
        10^{-7}}\right) \right)
 {1\over M_N^2} \overline{e}\gamma_i e\ \epsilon^{abi}
 \varepsilon^*_a\varepsilon_b
 \ \ \ ,
 \label{eq:pvmat}
\end{eqnarray}
where $ \varepsilon^*_a$ and $ \varepsilon_b$ are the polarization vectors of
the final and initial deuterons and $g_{\Delta S}$ is the strange axial 
matrix element of the nucleon.
Loop corrections in the standard model are fractionally larger 
than one would naively expect largely because of the suppression 
of the tree-level amplitude by a factor of 
$(1-4 \sin^2\theta_w)$\cite{MDDPKB}.
The estimated value of $h^{(1)}_{\pi NN}$ is in the range
$0\rightarrow 11\times 10^{-7}$ from the work of DDH\cite{DDH}, while
naive dimensional analysis estimates give $h^{(1)}_{\pi NN}
\sim 5\times 10^{-7}$\cite{KSa,DSLS}.
Present determinations of $g_{\Delta S}$ from lepton 
scattering\cite{SMC,EE}, including SU(3) breaking effects, 
are $\sim -0.15\pm 0.15$ (e.g. \cite{WSa}).
This suggests that the 
deuteron anapole moment contribution to ${\cal A}^{(pv)}$ is comparable to the 
contribution from $Z^0$ exchange.
However, for  values of $h^{(1)}_{\pi NN}$  suggested in \cite{PVprobs} or larger,  
the deuteron anapole moment  makes the larger contribution to ${\cal A}^{(pv)}$.
While it is clear that this will not provide the cleanest extraction
of $h^{(1)}_{\pi NN}$, 
we encourage the experimental community to continue their investigation of
the parity violating  
spin dependent interactions of the deuteron (e.g., \cite{sample})
in order to constrain  $h^{(1)}_{\pi NN}$ 
and help resolve the current controversy surrounding this parameter.
(Of course, if $h^{(1)}_{\pi NN}$ turns out to be much smaller than NDA
suggests, then the hadronic contribution we have computed here will
not be the dominant one.)

\section{Conclusions}

We have shown how to systematically include hadronic parity violation 
from the standard
model of electroweak interactions into an effective field theory description
of nucleon nucleon interactions, including photons and pions.
Exchange of a single potential pion with coupling $h^{(1)}_{\pi NN}$ is
the leading contribution to the weak scattering of two nucleons in the 
effective field theory expansion.
The most general parity violating 
four nucleon operators categorized by partial wave and  
isospin change were discussed and 
the lowest dimension operators  presented.
The leading four nucleon operator that violates parity has at least one  
pion in addition to nucleons.
As an example we computed the leading contributions to the 
deuteron anapole moment, which could be used to place a constraint on 
$h^{(1)}_{\pi NN}$.

\vskip 1in

We would like to thank Ernest Henley and
Michael Ramsey-Musolf for useful discussions.
This work is supported in part by the U.S. Dept. of Energy under
Grants No. DE-FG03-97ER4014 and DE-FG02-96ER40945.

\vfill\eject

\section{Erratum and Addendum}

Recent work by Khriplovich and Korkin (KK)\cite{KK} has demonstrated 
that a leading order (LO) contribution to the deuteron anapole moment
was omitted in our paper\cite{SScrap}.
In addition to this omission there is a factor of $-2$ error in one of the 
contributions.
In this ``Erratum and Addendum'' we correct the errors, and discuss the chiral
limit of this observable in more detail.

There is a contribution to the deuteron anapole moment arising 
from the isovector  magnetic moment of the nucleon which was 
omitted in \cite{SScrap}, as detailed in \cite{KK}.
This interaction is described by the lagrange density
\begin{eqnarray}
{\cal L}_{1,B} & = &
{e\over 2 M_N} N^\dagger
\left( \kappa_0 + \kappa_1 \tau_3 \right) {\bf \sigma} \cdot {\bf B} N
\ \ \ \ ,
\label{eq:NucmagX}
\end{eqnarray}
where
$\kappa_0 = {1\over 2} (\kappa_p + \kappa_n)$ and
$\kappa_1 = {1\over 2} (\kappa_p - \kappa_n)$
are the isoscalar and isovector nucleon magnetic moments in nuclear magnetons, 
with
\begin{eqnarray}
\kappa_p & =&  2.79285\ ,\qquad\kappa_n = - 1.91304
\ \ \ \ .
\label{eq:nucmagdefX}
\end{eqnarray}
The magnetic field  is conventionally defined 
${\bf B} ={\bf \nabla} \times {\bf A}$.
The experimentally determined values of $\kappa_0$ and $\kappa_1$ include
the pion loop corrections that we need to this order.

Using the same definition of the deuteron anapole moment as given in
\cite{SScrap} of
\begin{eqnarray}
  {\cal L} & = & i A_D\ {1\over M_N^2}\
  \epsilon_{abc} D^{a\dagger} D^b  \partial_\mu F^{\mu c}
  \ \ \ ,
\end{eqnarray}
our corrected result becomes 
\begin{eqnarray}
	A_D & = & -{e g_A h^{(1)}_{\pi NN} M_N^2 \over 12 \pi f}
\left[ \kappa_1\ { m_\pi+\gamma\over (m_\pi+2\gamma)^2}
\ +\ {1\over 2 m_\pi} 
\ -\  { m_\pi^2+3 m_\pi\gamma + 12\gamma^2\over 6 m_\pi (m_\pi+2\gamma)^2}
\right] 
\ \ \ ,
\label{eq:anaaX}
\end{eqnarray}
where $\gamma=\sqrt{M_N B}$ with $B$ the deuteron binding energy.
The first term in eq.~(\ref{eq:anaaX}) is generated by  the 
nucleon isovector magnetic moment.
The second term is the leading contribution from the nucleon anapole moment,
which is purely isoscalar.
The third term in eq.~(\ref{eq:anaaX}) arises from the pion exchange 
interaction between the nucleons.
Combining the last two terms gives the somewhat more compact expression,
\begin{eqnarray}
	A_D & = & -{e g_A h^{(1)}_{\pi NN} M_N^2 \over 12 \pi f }
\left[ \kappa_1\ { m_\pi+\gamma\over (m_\pi+2\gamma)^2}
\ +\  { 2 m_\pi + 9 \gamma \over 6 (m_\pi+2\gamma)^2}
\right] 
\ \ \ .
\label{eq:anabX}
\end{eqnarray}
Inserting the appropriate values for constants into 
eq.~(\ref{eq:anabX}), the tree-level parity violating matrix element 
for electron-deuteron scattering that
depends upon the deuteron spin becomes
\begin{eqnarray}
  {\cal A}^{(pv)} & = &
  10^{-7} \left( 5.4  g_{\Delta S} \ -\ 0.21 \left( {h^{(1)}_{\pi NN}\over
        10^{-7}}\right) \right)
 {1\over M_N^2} \ \epsilon^{abi}\ 
 \overline{e}\gamma_i e\ 
 \varepsilon^*_a\ \varepsilon_b
 \ \ \ .
 \label{eq:pvmatX}
\end{eqnarray}
This expression replaces eq.~(4.5) in \cite{SScrap}.
The coefficient of $h^{(1)}_{\pi NN}$ has been changed from $0.17$
to $0.21$.
The individual contributions to the numerical value of $0.21$ that appears 
in eq.~(\ref{eq:pvmatX}) are $0.17$ from the nucleon isovector magnetic moment,
$0.07$ from the anapole moment of the nucleon, and $-0.03$ from 
minimal coupling interactions
between the nucleons.
The amplitude is dominated by the interaction with the nucleon magnetic moment.
It is important to note that the expression given in 
eq.~(\ref{eq:pvmatX}) is valid only for momentum transfer
$|{\bf k}|\ll \gamma , m_\pi$.
For momentum transfers larger than this, there will be an additional 
form factor that multiples the coefficient of $h^{(1)}_{\pi NN}$.

In the limit of the deuteron having zero binding energy, $\gamma\rightarrow 0$,
$A_D$ becomes
\begin{eqnarray}
	A_D & = & -{e g_A h^{(1)}_{\pi NN} M_N^2 \over 12 \pi f m_\pi}
\left[ \kappa_1\ +\ {1\over 3}
\right] 
\ \ \ ,
\label{eq:analimaX}
\end{eqnarray}
which agrees with the analogous limit of the expression given in
\cite{KK}.
The expressions given in \cite{KK} are only valid in the limit that the 
deuteron $\gamma\ll m_\pi$, as is stated below eq.~(12) in \cite{KK}.
The approximation they make is justified for the actual numerical values 
of $\gamma$ and $m_\pi$.

A nontrivial check of our calculation can be made by considering the 
chiral limit $m_\pi\ll\gamma\ll\Lambda_\chi$.
Since the deuteron is an isoscalar object, with a 
vanishing one-pion strong coupling,  there cannot be a $1/m_\pi$ 
divergence in this limit.  
Here the charged pions cannot resolve the deuteron as a bound state 
of two nucleons and a logarithmic dependence on $m_\pi$ 
is the most divergent behavior that can appear.
For  $m_\pi\ll\gamma$  eq.~(\ref{eq:anabX}) becomes
\begin{eqnarray}
	A_D & = & -{e g_A h^{(1)}_{\pi NN} M_N^2 \over 48 \pi f \gamma}
\left[ \kappa_1\ +\ {3\over 2}
\right] 
\ \ \ ,
\label{eq:analimbX}
\end{eqnarray}
which has the expected form.
The behavior shown in eq.~(\ref{eq:analimbX})
does not agree with the results obtained in \cite{KK}, who find
\begin{eqnarray}
A_D & \sim & \left( \kappa_1-{7\over 24}\right){1\over\gamma} + {1\over m_\pi}
\ \ \ .
\label{eq:analimbKKX}
\end{eqnarray}
In fact, since it was assumed that $\gamma\ll m_\pi$ in deriving the 
results of \cite{KK}, the $m_\pi\ll\gamma$ limit cannot be taken, and 
the dependence on $\gamma/m_\pi$ is correct
only in the limit of small $\gamma/m_\pi$.

Our calculation does not include
contributions beyond leading order in $1/M_N$ or $1/\Lambda_\chi$
in the expressions for either the nucleon or deuteron anapole moment.
Beyond the terms proportional to $h^{(1)}_{\pi NN}$ that contribute
at higher orders,  as discussed in \cite{KK}, there are also terms
generated from additional weak interaction vertices 
involving pions and nucleons.
A detailed discussion of such interactions and their contribution to the 
nucleon anapole moment is given in \cite{KSa}.
Therefore, the higher order correction included in \cite{KK} 
is only one of several contributions, 
and only provides an indication of the size of higher order terms.

\vfill\eject

\end{document}